\documentclass[prl, aps, 10pt, showpacs, superscriptaddress,
               twocolumn, floatfix]{revtex4-1}
\usepackage{times,graphicx,color}
\begin{document}
\title{Search for light gauge bosons of the dark sector at MAMI}
\newcommand{\mainz}{\affiliation{
    Institut f\"{u}r Kernphysik,
    Johannes Gutenberg-Universit\"{a}t Mainz, 
    D-55099 Mainz, Germany}}
\newcommand{\clermont}{\affiliation{
Clermont Universit\'e, Universit\'e Blaise Pascal, CNRS/IN2P3, 
LPC, BP 10448, F-63000 Clermont-Ferrand, France}}
\newcommand{\zagreb}{\affiliation{
    Department of Physics, 
    University of Zagreb, 
    HR-10002 Zagreb, Croatia}}
\newcommand{\stefan}{\affiliation{
    Jo\v{z}ef Stefan Institute, 
    SI-1000 Ljubljana, Slovenia}}
\newcommand{\ljubljana}{\affiliation{
    Department of Physics, 
    University of Ljubljana, 
    SI-1000 Ljubljana, Slovenia}}  
\author{H.~Merkel}
\thanks{merkel@kph.uni-mainz.de}
\mainz
\homepage{http://www.kph.uni-mainz.de}
\author{P.~Achenbach}
\author{C.~Ayerbe~Gayoso}
\author{J.~C.~Bernauer}
\thanks{Present address: MIT-LNS, Cambridge, MA, USA.}
\author{R.~B\"{o}hm}
\mainz
\author{D.~Bosnar}
\zagreb      
\author{L.~Debenjak}
\stefan
\author{A.~Denig}
\author{M.~O.~Distler}
\author{A.~Esser}
\mainz
\author{H.~Fonvieille}
\clermont
\author{I.~Fri\v{s}\v{c}i\'{c}}
\zagreb
\author{D.~G.~Middleton}
\author{U.~M\"{u}ller}
\author{L.~Nungesser}
\author{J.~Pochodzalla}
\author{M.~Rohrbeck}
\author{S.~S\'{a}nchez Majos}
\author{B.~S.~Schlimme}
\author{M.~Schoth}
\mainz   
\author{S.~\v{S}irca}
\stefan
\ljubljana
\author{M.~Weinriefer}
\mainz
\collaboration{A1 Collaboration}
\date{January 21, 2011}
\noaffiliation
\begin{abstract}
  A new exclusion limit for the electromagnetic production of a light
  $U(1)$ gauge boson $\gamma'$ decaying to ${e^+e^-}$ was
  determined by the A1 Collaboration at the Mainz Microtron. Such
  light gauge bosons appear in several extensions of the standard
  model and are also discussed as candidates for the interaction of
  dark matter with standard model matter. In electron scattering from
  a heavy nucleus, the existing limits for a narrow state coupling to
  ${e^+e^-}$ were reduced by nearly an order of magnitude in
  the range of the lepton pair mass of $210\,\mathrm{MeV}/c^2 <
  m_{{e^+e^-}} < 300\,\mathrm{MeV}/c^2$. This experiment
  demonstrates the potential of high current and high resolution fixed
  target experiments for the search for physics beyond the standard
  model.
\end{abstract}
\maketitle

\paragraph{Introduction.---\hspace{-3mm}}

An additional $U(1)$ interaction appears to be natural in nearly all
theoretical extensions of the standard model. Large gauge symmetries
have to be broken and $U(1)$ bosons provide the lowest-rank local
symmetries. For example, in standard embedding of most variants of
string theories a $U(1)$ boson is generated by symmetry breaking.
Such additional $U(1)$ bosons may be hidden; e.g., no
standard model particles are charged under the corresponding symmetry,
but their mass is allowed in the range of the standard model masses.

Recently, several experimental anomalies were discussed as possible
signatures for a hidden force. A light $U(1$) boson in the mass range
below $1\,\mathrm{GeV}/c^2$ might explain e.g. the observed
anomaly of the muon magnetic moment
\cite{PhysRevD.73.072003,Pospelov:2008zw}. Cosmology and astrophysics
provide an abundant amount of evidence for the existence of dark
matter (for a summary, see, e.g., Ref.~\cite{PDG}). Several
experimental hints point to a $U(1)$ boson coupling to leptons as the
mediator of the interaction of dark matter with standard model matter
(see, \textit{e.g.}, Ref.~\cite{ArkaniHamed:2008qn} for a detailed
discussion).  For example the lively debated annual modulation signal
of the DAMA-LIBRA experiment \cite{Bernabei:2008yi} could be brought
into accordance with the null result of bolometric experiments if one
assumes an interaction via a light $U(1)$ boson
\cite{Borodatchenkova:2005ct}. Observations of cosmic rays show a positron
excess \cite{Adriani:2008zr}. While this excess may be due to
astrophysical process like quasars, this could also be a hint for the
annihilation of dark matter into leptons. If the experimental evidence
is interpreted as annihilation of dark matter, the excess of positrons
and no excess of antiprotons in cosmic rays hints again to a mass of
the $U(1)$ boson below $2\,\mathrm{GeV}/c^2$.

The interaction strength of such a $U(1)$ boson (in the following
denoted as $\gamma'$, in the literature also denoted as $A'$, $U$, or
$\phi$) with standard model particles is governed by the mechanism of
kinetic mixing \cite{Holdom:1985ag}. The coupling can be subsumed by
an effective coupling constant $\epsilon$ and a vertex structure of a
massive photon.

\begin{figure}[b]
\includegraphics[width=0.9\columnwidth]{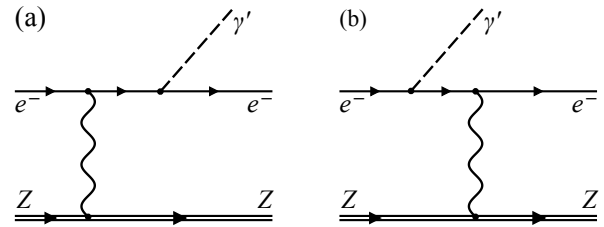}
\caption{Electromagnetic production of the $\gamma'$ boson. The
  coupling of the $\gamma'$ boson is parametrized as $i\,\epsilon\,e\,
  \gamma^\mu$.}
\label{signal}
\end{figure}

Bjorken \textit{et al.}\ \cite{Bjorken:2009mm} discussed several
possible experimental schemes for the search of a $\gamma'$ in the
most likely mass range of a few $\mathrm{MeV}/c^2$ up to a few
$\mathrm{GeV}/c^2$. Since the coupling is small, the cross section for
coherent electro magnetic production of the $\gamma'$ boson can be
enhanced by a factor $Z^2$ by choosing a heavy nucleus as the target
(see Fig.~\ref{signal}). The subsequent decay of the $\gamma'$ boson
to a lepton pair is the signature of the reaction.

The cross section of signal and background were estimated in
Ref.~\cite{Bjorken:2009mm} in the
Weizs\"acker-Williams approximation. In this approximation, the cross
section shows a sharp peak, in both signal and background, where
nearly all the energy of the incident electron is transferred to the
lepton pair $(E_{e^+}+E_{e^-})=E_0$. Correspondingly,
the pair is produced dominantly in the direction of the incident
electron.

\begin{table*}
  \parbox{0.7\textwidth}{
    \caption{Kinematic settings. The incident beam energy was $E_0 =
      855\,\mathrm{MeV}$, and the settings are roughly centered around
      $E_{e^+}+E_{e^-}=E_0$ and
      $m_{\gamma'}=250\,\mathrm{MeV}/c^2$.}}
  \begin{tabular*}{0.7\textwidth}{@{\extracolsep{\fill}}lrrcrrcr}
    \hline\hline
    & \multicolumn{3}{c}{Spec. A (${e}^+$) }
    & \multicolumn{3}{c}{Spec. B (${e}^-$) }\\
    & \multicolumn{1}{c}{$p$ (MeV)} & \multicolumn{1}{c}{$\theta$}
    & \multicolumn{1}{c}{$d\Omega$ (msr)} 
    & \multicolumn{1}{c}{$p$ (MeV)} & \multicolumn{1}{c}{$\theta$}
    & \multicolumn{1}{c}{$d\Omega$ (msr)}&\multicolumn{1}{c}{Events}\\
    \hline 
    Set-up 1 & 346.3 & $22.8^\circ$ & 21 & 507.9 & $15.2^\circ$ & 5.6 &$208\times 10^6$\\
    Set-up 2 & 338.0 & $22.8^\circ$ & 21 & 469.9 & $15.2^\circ$ & 5.6 &$47\times 10^6$\\
    \hline\hline
  \end{tabular*}
  \label{tab}
\end{table*}

\begin{figure}
  \includegraphics[width=0.9\columnwidth]{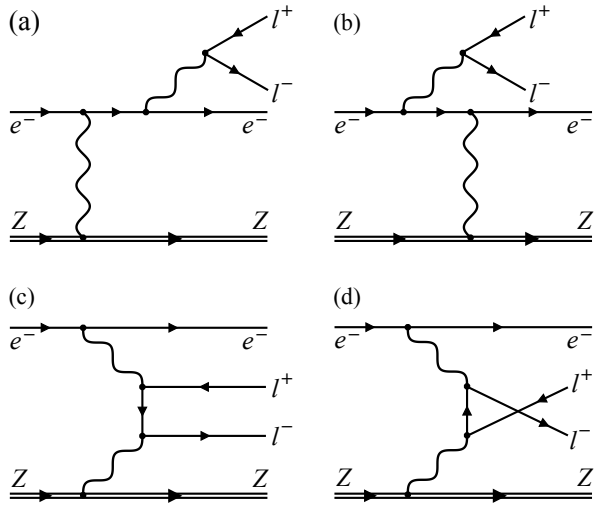}
  \caption{Dominant background processes. While graphs (a) and (b)
    have the same structure as the signal and present an irreducible
    background, the contributions of graphs (c) and (d) can be
    suppressed by the choice of kinematic setting.}
  \label{background}
\end{figure}

The experimental challenge is the suppression of the background which
is dominated by radiative pair production
(Fig.~\ref{background}). Radiation by the final or initial electron
[Figs.~\ref{background}(a) and ~\ref{background}(b)] has the same
cross-section structure as the desired signal and is an irreducible
background to this experiment. Radiation with an internal lepton line
[Figs.~\ref{background}(c) and ~\ref{background}(d)] has a maximum if the
internal electron line is nearly on the mass shell, i.e., if one of the
leptons carries nearly all the energy of the pair. This background can
be reduced by choosing a kinematic setting in which the electron and
positron are detected at equal angles and momenta.

\paragraph{Experiment.---\hspace{-3mm}}

The experiment took place at the spectrometer setup of the A1
Collaboration at the Mainz Microtron (MAMI) (see
Ref.~\cite{Blomqvist:1998xn} for a detailed description). An
unpolarized electron beam with a beam energy of
$E_0=855\,\mathrm{MeV}$ and a beam current of $90\,\mu\mathrm{A}$ was
incident on a tantalum foil (99.9\% $^{181}$Ta, $Z=73$) with an area
density of $81.3\,\mathrm{mg/cm^2}$, leading to a luminosity of
$L\,Z^2=8.07\cdot10^{38}\,\mathrm{s^{-1}cm^{-2}}$. The beam was
rastered across the target to reduce the local thermal load on the
target foil.

\begin{figure}
  \includegraphics[width=\columnwidth]{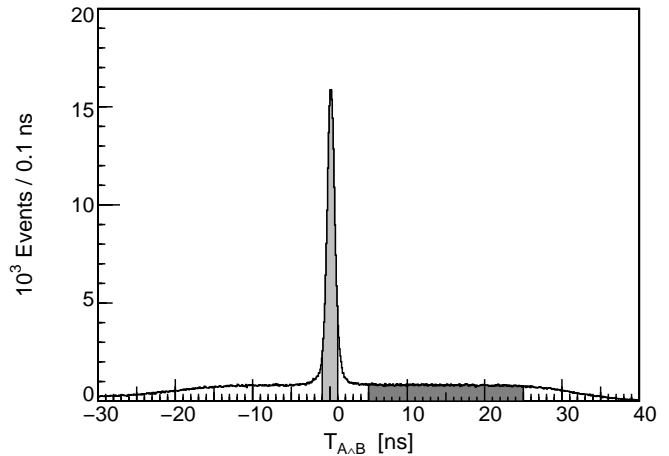}
  \caption{Coincidence time distribution after particle identification
    by \v{C}erenkov detectors (set-up 1). The events of the light
    shaded area were used as true coincidences, while the dark shaded
    area was used as an estimate of the accidental coincidences.}
  \label{timing}
\end{figure}

\begin{figure}
  \includegraphics[width=\columnwidth]{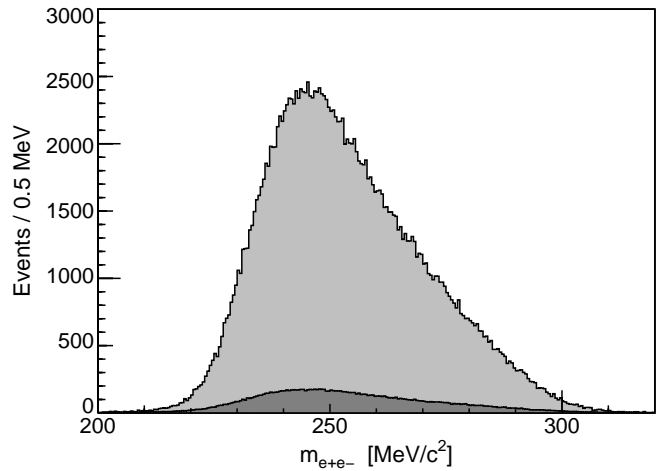}
  \caption{Mass distribution of the reconstructed $e^+$-$e^-$ pair
    (setup 1). The dark shaded area denotes the background due to
    accidental coincidences (scaled to a time window of 2\,ns).}
  \label{mass}
\end{figure}

\begin{figure*}
\includegraphics[width=\textwidth]{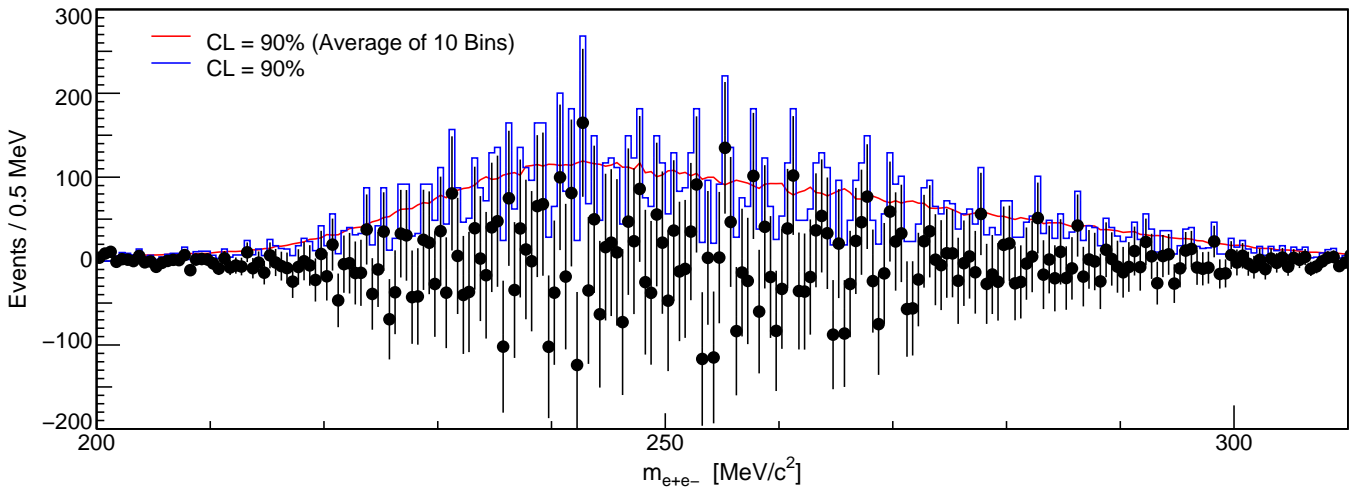}
\caption{Upper exclusion limits with 90\% confidence level determined
  by the Feldman-Cousins algorithm (all data). The
  averaged limit is included for subjective judgement only ($\approx
  10\%$ of the data points should be above this line at 90\% C.L.).}
\label{exclusion}
\end{figure*}

For the detection of the electron-positron pair, two high-resolution
spectrometers were used. The particles were detected by vertical drift
chambers for tracking and scintillator detectors for trigger and
timing purpose. In addition, a threshold-gas-\v{C}erenkov detector was
used in each arm to discriminate between electrons or positrons and
pions.


Table~\ref{tab} summarizes the kinematic setups used. Setup 1 was
chosen to be close to $E_{e^+}+E_{e^-}=E_0$ where the
cross section has a sharp peak to ensure high count rates. In
addition, setup 2 was selected at $E_{e^+} + E_{e^-} =
0.9\,E_0$ during the experiment to optimize the total count rates. The
angles of the spectrometers were set to be nearly symmetric to reduce
the background by the Bethe-Heitler process [Figs.~\ref{background}(c) and
\ref{background}(d)]. In total, data of four days of beam time were
used for the analysis. The electrons and the positrons were detected
by the coincidence of the raw scintillator signals. The \v{C}erenkov
signals were not included in the trigger logic but recorded for
off-line analysis.

\paragraph{Data analysis.---\hspace{-3mm}}

Only events with a positive signal in the \v{C}erenkov detectors were
selected with an efficiency of 98\% for spectrometer A and 95\% for
spectrometer B \cite{Blomqvist:1998xn}. Figure~\ref{timing} shows the
coincidence time between the corresponding spectrometers after
correction for the flight path of $\approx 12\,\mathrm{m}$ within the
spectrometers for these events. A timing resolution of better than
1\,ns FWHM was achieved, and a cut of
$-1\,\mathrm{ns}<t_\mathrm{A\wedge{}B} < 1\,\mathrm{ns}$ was used to
mark the true electron-positron pairs. Below the peak, a background due
to accidental coincidences is present. To estimate this background,
events in the coincidence side band $5\,\mathrm{ns} <
t_\mathrm{A\wedge B} < 25\,\mathrm{ns}$ were selected and weighted by
the ratio of the timing windows.

For the real electron-positron pairs the invariant mass squared of the
pair was determined by the four-momentum sum
$m_{e^+e^-}^2={\left(p_{e^+} +
    p_{e^-}\right)^2}$. Figure~\ref{mass} shows the resulting
mass spectrum. The contribution of the accidental background is indicated
by the dark shaded area.

In this figure, a possible candidate for the dark photon would appear
as a peak on top of the background. The width of such a peak can only
be estimated by simulation. For this, the experimental resolution of
the four-vector determination of a single spectrometer was determined
by the width of the lowest lines of the nuclear excitation spectrum in
elastic electron scattering. This single spectrometer resolution was
used as input for the simulation of the experiment. A mass resolution
of better than $0.5\,\mathrm{MeV}/c^2$ was determined, corresponding
to the chosen bin width in Fig.~\ref{mass}.

No significant peak in the mass spectrum was observed. The
corresponding upper limit was determined by the Feldman-Cousins
algorithm \cite{Feldman:1997qc}. As input for this algorithm the raw
mass spectrum was used, and as a background estimate for each bin the
mean of the three neighboring bins on either side was used. This
choice of the background estimate introduces systematic errors, which
have to be investigated in the case of a positive signal but only
enhance statistical fluctuations in the case of an upper
limit. Figure~\ref{exclusion} shows the resulting exclusion limits.

\paragraph{Results and Interpretation.---\hspace{-3mm}}

In order to interpret the result in terms of the effective coupling
$\epsilon$ of a possible dark photon candidate, a model for the
production process has to be used. Unfortunately, it turns out that
the Weizs\"{a}cker-Williams-approximation used in
Ref.~\cite{Bjorken:2009mm} fails in this energy range by orders of
magnitude, mainly since the recoil of the nucleus cannot be neglected.
Taking into account the nuclear recoil, the peak at
$(E_{e^+}+E_{e^-})=E_0$ in Ref.~\cite{Bjorken:2009mm} is
regularized and the cross section at this point becomes zero. In
addition, the assumption of a real initial photon exactly in the direction
of the electron beam introduces a peak in the angular distribution,
which is not present in electro production due to helicity
conservation of the scattered electron.

\begin{figure}
  \includegraphics[width=\columnwidth]{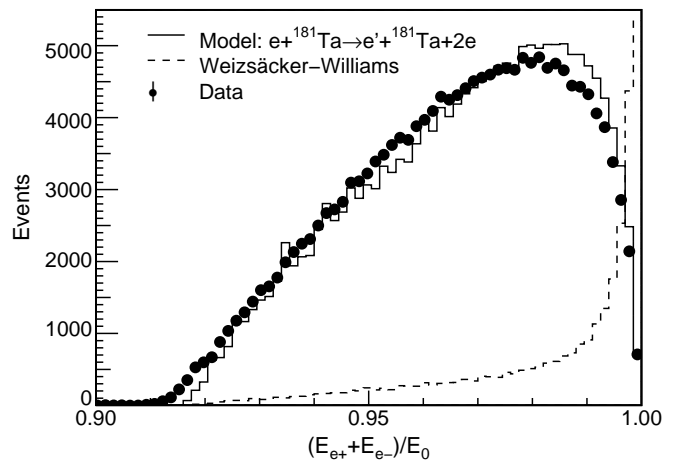}
  \caption{Comparison of simulation with data (setup 1). As a model the coherent
    electro-production from a heavy nucleus was used.}
  \label{simulation}
\end{figure}

Instead, we used as a model for the $\gamma'$ production the coherent
electro production from the tantalum nucleus, calculated as the
coherent sum of the graphs of Fig.~\ref{signal}. The charge
distribution of tantalum was approximated as a solid sphere. For the
QED background we used the coherent sum of the graphs of
Fig.~\ref{background}. The corresponding cross sections were included
on an event by event basis in the simulation. The simulation including
this model shows excellent agreement with data, as demonstrated in
Fig.~\ref{simulation}, where the background-subtracted yields as an
estimate for the QED background graphs are compared to the simulation
of this process.

The remaining model dependence of this interpretation mainly affects
the nuclear vertex, since, e.g., the possible breakup of the recoil
nucleus is neglected. Since this vertex is common to both the signal
and the QED background channels, to further reduce the model
dependence we use only the ratio of the signal to QED background of
the simulation in addition to the accidental background. The ratio can
be translated to the effective coupling for a given mass resolution
$\delta_m$ by using Eq.~(19) of Ref.~\cite{Bjorken:2009mm}:
\[
\frac{d\sigma({X\rightarrow \gamma'\,Y\rightarrow e^+e^- Y})}{
  d\sigma({X\rightarrow \gamma^* Y\rightarrow e^+e^- Y})} =
\frac{3\pi}{2N_\mathrm{\mathrm{f}}} \frac{\epsilon^2}{\alpha}
\frac{m_{\gamma'}}{\delta_m}
\]
and the measured event rate as an estimate for the background channel.
The number of final states $N_\mathrm{f}$ includes the ratio of phase
space for the corresponding decays above the $\mu^+\mu^-$ threshold.

\begin{figure}
  \includegraphics[width=\columnwidth]{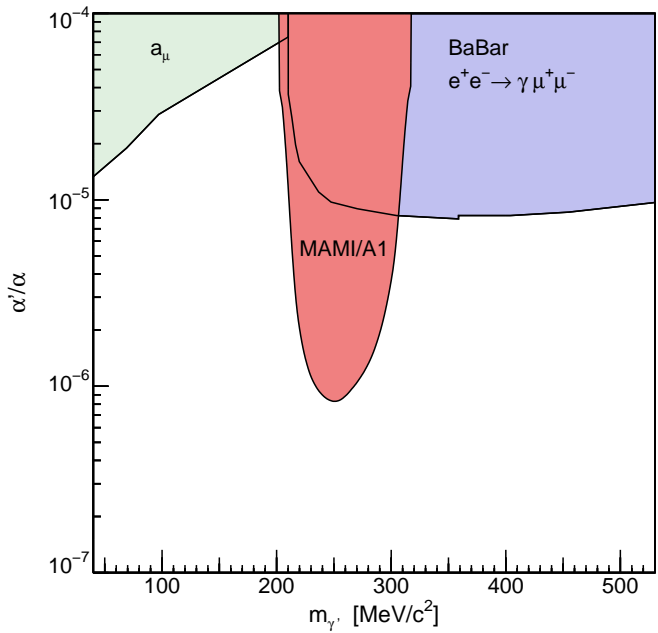}
  \caption{Exclusion limits with 90\% confidence level in terms of
    relative coupling $\alpha'/\alpha=\epsilon^2$. Also shown are the
    previous results by \textit{BABAR} \cite{Aubert:2009cp} and for
    $a_\mu$ of the muon \cite{Pospelov:2008zw}.}
  \label{exclusionall}
\end{figure}

Figure \ref{exclusionall} shows the result of this experiment in terms
of the ratio of the effective coupling to the fine structure constant
$\alpha'/\alpha = \epsilon^2$. For clarity of the figure, the
exclusion limit was averaged. Also shown are the existing limits
published by BaBar \cite{Aubert:2009cp} and the standard model
prediction \cite{Pospelov:2008zw} of the muon anomalous magnetic
moment $a_\mu =g_\mu/2-1$ (calculation of exclusion limits in
$\epsilon^2$ by \cite{Essig:2010xa}). The existing exclusion limit has
been extended by an order of magnitude.

In this experiment, the discovery potential of the existing high
luminosity electron accelerators has been demonstrated. The background
conditions are well under control due to excellent timing and missing
mass resolution. An extensive program to cover further mass regions
with similar experiments is planned at MAMI, Jefferson
Lab\cite{Essig:2010xa}, and other laboratories (for a review see Ref.
\cite{Andreas:2010tp}).

\acknowledgments{The authors thank the MAMI accelerator
  group for providing the excellent beam quality and intensity
  necessary for this experiment and T. Beranek for fruitful
  discussions on the QED calculations. This work was supported by the
  Federal State of Rhineland-Palatinate and by the Deutsche
  Forschungsgemeinschaft with the Collaborative Research Center 443.
}
\bibliographystyle{apsrev4-1} 
\bibliography{dm}
\end{document}